\documentclass[aps,prl,twocolumn]{revtex4-1}
\usepackage[utf8]{inputenc}
\usepackage[T1]{fontenc}
\usepackage[english]{babel}
\usepackage{amssymb,amsmath,mathtools} 
\usepackage{graphicx} 
\usepackage{xcolor}
\usepackage{bbold}
\usepackage{braket}
\usepackage[headings]{fullpage}
\usepackage{hyperref}
\usepackage{slashed}
\usepackage[titletoc,toc,title]{appendix}
\usepackage[hang,small,bf]{caption}
\usepackage[nottoc,numbib]{tocbibind}
\usepackage[export]{adjustbox}
\usepackage{xcolor}
\usepackage{array}
\usepackage{booktabs}
\usepackage{hyperref}
\usepackage{amsthm}

\newcommand{\re}{{\rm Re}\,}

\begin{document}
	\newcommand{\be}{\begin{eqnarray}}
		\newcommand{\ee}{\end{eqnarray}}
	\newcommand{\del}{\partial}
	\newcommand{\nn}{\nonumber}
	\newcommand{\STr}{{\rm Str}}
	\newcommand{\Sdet}{{\rm Sdet}}
	\newcommand{\Pf}{{\rm Pf}}
	\newcommand{\mat}{\left ( \begin{array}{cc}}
		\newcommand{\emat}{\end{array} \right )}
	\newcommand{\vect}{\left ( \begin{array}{c}}
		\newcommand{\evect}{\end{array} \right )}
	\newcommand{\tr}{{\rm Tr}}
	\newcommand{\hm}{\hat m}
	\newcommand{\ha}{\hat a}
	\newcommand{\hz}{\hat z}
	\newcommand{\hze}{\hat \zeta}
	\newcommand{\hx}{\hat x}
	\newcommand{\hy}{\hat y}
	\newcommand{\tm}{\tilde{m}}
	\newcommand{\ta}{\tilde{a}}
	\newcommand{\U}{\rm U}
	\newcommand{\D}{\slashed{D}}
	\newcommand{\hc}{^\dagger}
	\newcommand{\inv}{^{-1}}
	\newcommand{\diag}{{\rm diag}}
	\newcommand{\sign}{{\rm sign}}
	\newcommand{\ct}{\tilde{c}}
	\newtheorem{theorem}{Conjecture}

\title{On the Role of Spatial Effects in Early Estimates of Disease Infectiousness:\\A Second Quantization Approach}

\author{Adam Mielke}\email{admi@dtu.dk}\affiliation{Dynamical Systems, Technical University of Denmark, Asmussens Allé, 303B, 2800 Kgs.\ Lyngby, Denmark}

\date{\today}
 
\date{\today}
\maketitle

\section{Summary}
With the covid-19 pandemic still ongoing and an enormous amount of test data available, the lessons learned over the last two years need to be developed to a point where they can provide understanding for tackling new variants and future diseases.

The SIR-model \cite{SIR1, SIR2, SIR3}, commonly used to model disease spread \cite{SpatialEpidemicsPoland, SpatialEpidemicsStatMech, EkspertgruppenRapporter, SpatialEpidemicsGraph, SpatialEpidemicsFluid1, SpatialEpidemicsFluid2, SpatialEpidemicsFluid3, Liu, SIR4, SIR5}, predicts exponential initial growth, which helps establish the infectiousness of a disease in the early days of an outbreak.

Unfortunately, the exponential growth becomes muddied by spatial, finite-size, and non-equilibrium effects in realistic systems \cite{SpatialEpidemicsChallenge, SpatialEpidemicsPoland, SpatialEpidemicsStatMech, Liu, SpatialEpidemicsFluid1, SpatialEpidemicsFluid2, SpatialEpidemicsGraph, SpatialEpidemicsFluid3, SpatialEpidemicsData1, SpatialEpidemicsData2}, and robust estimates that may be used in prediction and description are still lacking.

I here establish a second quantization framework that allows introduction of arbitrarily complicated spatial behavior, and I show that a simplified version of this model is in good agreement with both the growth of different covid-19 variants in Denmark, a simulation study, and analytical results from the theory of branched polymers. Denmark is well-suited for comparison, because the number of tests with variant information in early December 2021 is very high, so the spread of a single variant can be followed.

I expect this model to build bridges between the epidemic modeling and solid state communities. The long-term goal of the particular analysis in this paper is to establish priors that allow better early estimates for the infectiousness of a new disease.

\subsection{Article Structure}
The main article is divided into two sections. The framework and model description come first, and then comes comparison of the spatial corrections to the growth of different covid-19 variants.
The first part assumes a basic familiarity with quantum operators, but if the result in Equation \eqref{Eq:NFinal} is accepted as true, the second part, including the model interpretation, can be read independently.

\section{Model Description and Solution}
Let me introduce a model in terms of creation ($c\hc,b\hc$) and annihilation ($c,b$) operators. These are well-established in many areas of quantum physics \cite{Srednicki, BruusFlensberg} and have also seen some use in disease modeling \cite{SIR-2ndQuant1, SIR-2ndQuant2, SIR-2ndQuant3, SIR-2ndQuant4}. But whereas previous papers have focused on a reformulation of compartmental models in terms of these operators, I will establish two full Hamiltonians, one fermionic and one bosonic, both of the same form
\begin{eqnarray}\label{Eq:Hamiltonians}
	H_{\rm fer} &=& \beta \sum_{jk} c\hc_j A_{jk} N_k + \gamma\sum_j c_j\\
	&& + \frac{1}{2}\left(\mu\sum_{jk} c\hc_j B_{jk} c_k + h.c.\right)\nn\ ,\\
	H_{\rm bos} &=& \beta \sum_{jk} b\hc_j A_{jk} N_k + \gamma\sum_j b_j N_j\\
	&&+ \frac{1}{2}\left(\mu\sum_{jk} b\hc_j B_{jk} b_k + h.c.\right)\ ,\nn
\end{eqnarray}
which both have spatial structure built in directly. The operators $c\hc$ and $b\hc$ are fermionic and bosonic creation operators respectively at site $j$, and $N_j$ is the counting operator with eigenvalue $n_j$. (For the fermionic model, $n_j = 0, 1$, which is also why the counting operator is redundant in the $\gamma$-term.)

The epidemiological interpretation is that each site is a person (or household), and creating a particle at a site corresponds to infecting that person.

The matrix entries $A_{jk}$ allow a particle to be created at site $j$ if there already is one at site $k$. In this sense, it mimics the spread of disease along the pathways determined by the $A$. The $\gamma$-term allows recovery from disease. These two terms do not conserve particle number, and so the Hamiltonians are non-Hermitian. The $B$-term is responsible for spread of infection without particle generation.

The coefficient notation is borrowed from SIR-models, but the translation between the two turns out to be non-trivial. Although the physical interpretations of the two Hamiltonians are very different, they may be treated simultaneously because of the similar commutation relations.
For full generality I bound each site $j$ of the bosonic Hamiltonian at $\nu_j$ particles. These may be taken to infinity if needed.

Of course there have been numerous works on the spatial epidemics. The main challenges were outlined in \cite{SpatialEpidemicsChallenge}, and more recent works provide models that describe different aspects. These either focus on a fluid dynamical description \cite{SpatialEpidemicsFluid1, SpatialEpidemicsFluid2, SpatialEpidemicsFluid3}, network dynamics \cite{SpatialEpidemicsStatMech, Liu, SpatialEpidemicsPoland, SpatialEpidemicsGraph}, or on data analysis \cite{SpatialEpidemicsData1, SpatialEpidemicsData2}. An analytical predictions for the growth is still missing, which is the goal of this section.

To study growth rates, it becomes necessary to look how the number of particles (i.e., the amount of infection) behaves as a function of time
\begin{eqnarray}
\sum_j \bra{\psi(t)}N_j\ket{\psi(t)} &=& \sum_j \bra{\psi(0)}e^{iH\hc t^*}N_je^{-iHt}\ket{\psi(0)}\nn\\
\end{eqnarray}
under the initial condition
\begin{eqnarray}
	\ket{\psi(0)}_{\rm fer} = c_a\hc \ket{0} \ &,&\ \ket{\psi(0)}_{\rm bos} = b_a\hc \ket{0}\ .
\end{eqnarray}
This will be the definition of the index $a$. To see the connection to statistical models, I will make a Wick rotation $t = i\tau$. This is a well-established method \cite{Srednicki, SIR-2ndQuant1, SIR-2ndQuant2, BruusFlensberg} and makes the time evolution
\begin{eqnarray}
	\ket{\psi(\tau)} = e^{H\tau}\ket{\psi(0)}\ &,&\ \bra{\psi(\tau)} = \bra{\psi(0)}e^{H\hc\tau^*}\ .
\end{eqnarray}
Note that I here diverge from convention, see for example \cite[Chapter 9]{BruusFlensberg}, where the Wick-rotated time evolution of an operator $\mathcal{O}$ is written as $\mathcal{O}(\tau) = e^{H\tau} \mathcal{O}(0)e^{-H\tau}$ for a Hermitian Hamilton (i.e.\ ignoring the complex conjugation of time in $\bra{\psi(t)}$). This makes chaining time evolutions simpler, but breaks Hermiticity and thus does not guarantee a real observable. Interpreting the Wick rotation as a theory in Euclidian space rather than Minkowskian, the observables must be physical, and I therefore use this sign convention. This is closer to the high-energy physics interpretation \cite{Srednicki}.
I want to emphasize that $\tau$ real is the relevant model here, that is, where time has simply been rotated $\frac{\pi}{2}$ in the complex plane. However, for transparency I continue to treat $\tau$ like a complex parameter.

The problem is thus reduced to one of commutation relations, and because of the similar structure
\begin{eqnarray}
	\left[N_j,\ b_k\right] = -\delta_{jk}b_j\ &,&\ \left[N_j,\ b\hc_k\right] = \delta_{jk}b\hc_j\label{Eq:Bos:Com}\\
	\left[N_j,\ c_k\right] = -\delta_{jk}c_j\ &,&\ \left[N_j,\ c\hc_k\right] = \delta_{jk}c\hc_j\label{Eq:Fer:Com}\ ,
\end{eqnarray}
I can treat the two Hamiltonians simultaneously. As I never mix bosonic and fermionic states, $N_j$ will simply denote the counting operator of the corresponding model.

Consider the special case $\gamma = 0$ and $B_{jk} = \delta_{jk}$. That is,
\begin{eqnarray}
	H_{\rm fer}  &=& \beta \sum_{jk} c\hc_k A_{kj} N_j + \re(\mu)\sum_j N_j\nn\\
	&\equiv& H_0 + \re(\mu)\sum_j N_j \ .
\end{eqnarray}
and the same for the bosonic model.
This is simple enough to calculate because the commutator is recursive
\begin{eqnarray}
	\left[\re(\mu)\sum_j N_j,\ H_0\right] &=& \re(\mu)H_0\ .
\end{eqnarray}
This allows me to adapt a special case of the Baker-Campbell-Hausdorff formula \cite{HallHandbook} to
\begin{eqnarray}
	e^{\tau (H_0 + \re(\mu)\sum_j N_j)} &=& e^{\tau\frac{e^{\re(\mu)} - 1}{\re(\mu)}H_0} e^{\tau\re(\mu)\sum_j N_j}\nn\\
\end{eqnarray}
This is significant because it means the counting operator can be applied first and independently of $H_0$. Because the initial state is an eigenstate of the counting operator, the following is obtained
\begin{eqnarray}\label{Eq:NFinal}
	N(\tau) &=& e^{2\re(\mu)\re(\tau)}\\
	&&\times\sum_{\rho=0}^{\infty} \frac{\left(1 + \rho\right)\left(\frac{e^{\re(\mu)} - 1}{\re(\mu)}|\beta| |\tau|\right)^{2\rho}}{(\rho!)^2} ||A||^{\rho, a}\ ,\nn
\end{eqnarray}
for $N(0)=1$. It is very natural that Equation \eqref{Eq:NFinal} is an even function of time. Firstly, because particle number is invariant under time reversal $t\to -t$ for real time $t$. Secondly, the first two terms may be interpreted as contributions from random walks and branched polymers, which have extrinsic Hausdorff dimension 2 and 4 respectively \cite{Ambjoern}, regardless of the embedded space. (The intrinsic dimension is 2 for both branched polymers and random walk.)
Detailed derivation of this Equation \eqref{Eq:NFinal} as well as the definition of the norms $||A||^{\rho, a}$ may be found in the supplementary material.

\section{Interpretation and Comparison to the Covid-19 Epidemic}
The result \eqref{Eq:NFinal} has two parts: One purely exponential, determined by $\mu$, and one more complicated function, which here has been written as an expansion with coefficients depending on $\beta A$.

\begin{figure*}
	\centering
	\begin{tabular}{cc}
		\hspace{20pt}EU2020 variants & \hspace{20pt}Omicron variant\\
		\includegraphics[width=0.3\textwidth]{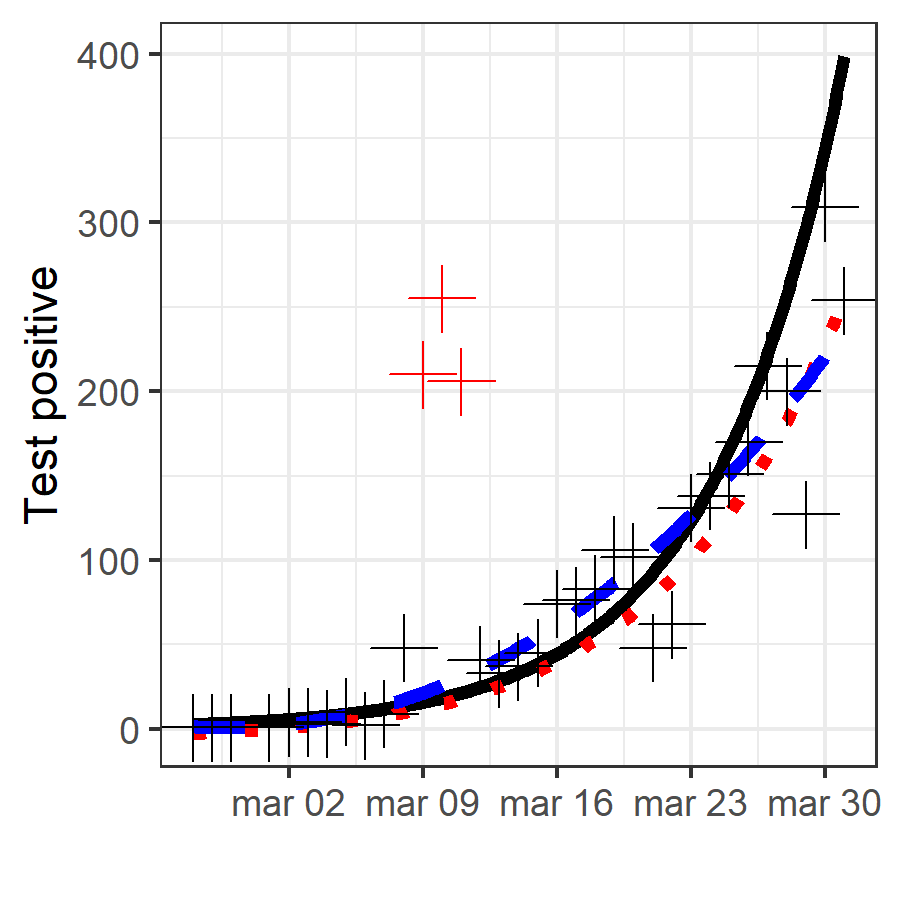} &
		\includegraphics[width=0.3\textwidth]{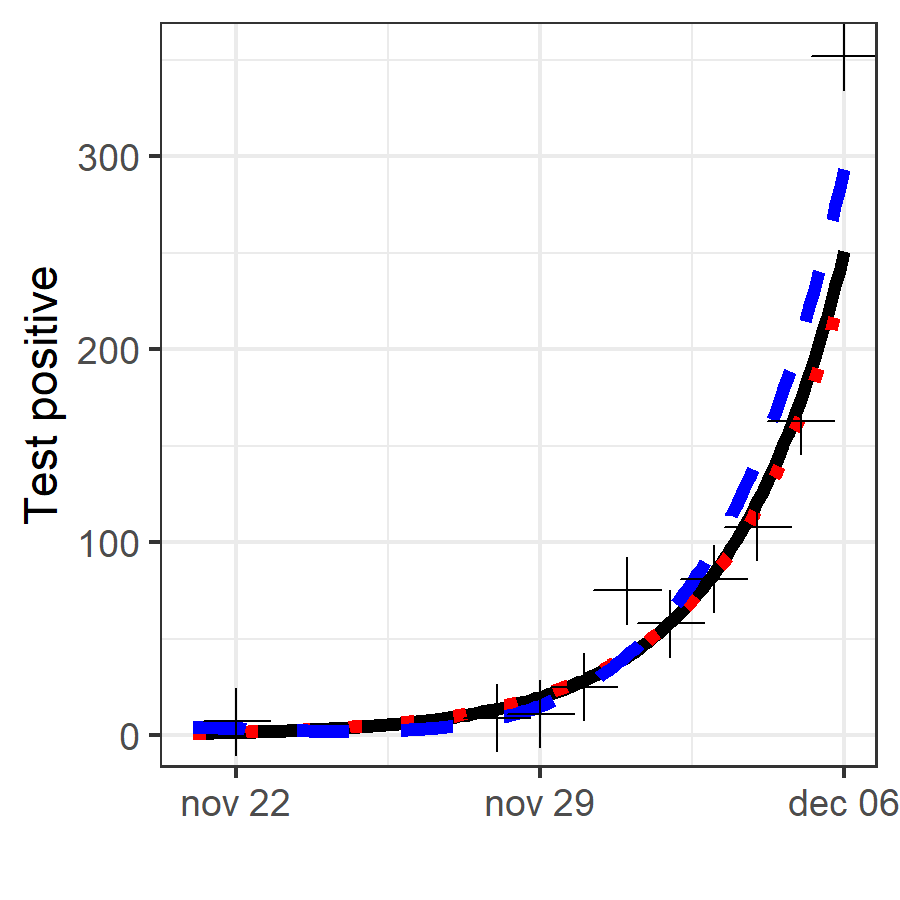}\\
		\includegraphics[width=0.3\textwidth]{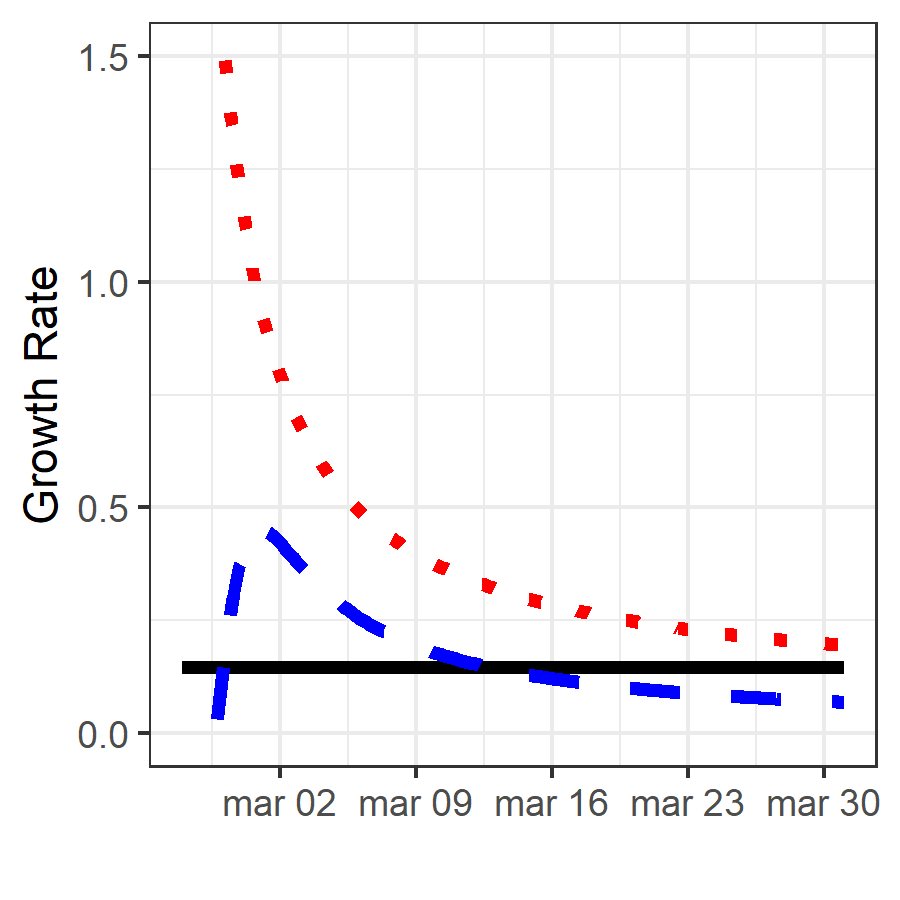} &
		\includegraphics[width=0.3\textwidth]{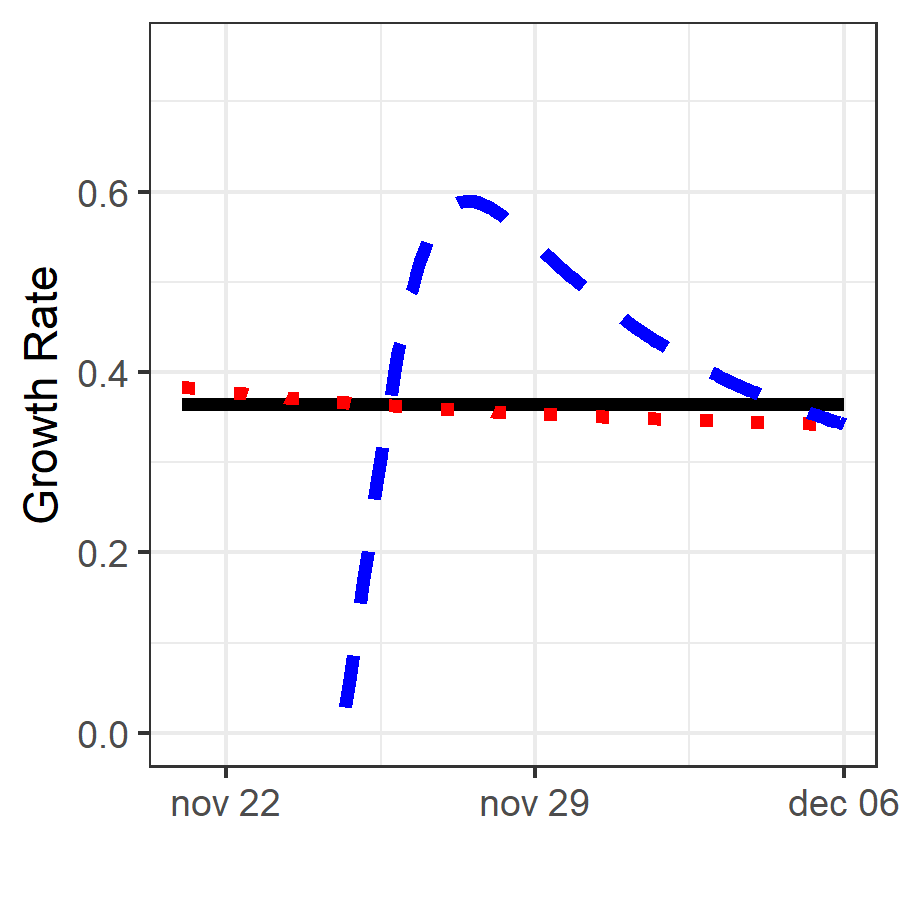}
	\end{tabular}
	\caption{Comparison to the covid-19 epidemic in Denmark using the three models: No correction (Equation \eqref{Eq:ThreeModels1}, black solid line), linear correction (Equation \eqref{Eq:ThreeModels2}, red dotted line), and quadratic correction (\eqref{Eq:ThreeModels3}, blue dashed line). Both the emergence of the EU2020-strains in the spring of 2020 \cite{SSIdata} (left column) and the introduction of the omicron variant in the late fall of 2021 \cite{EkspertgruppenMatricer} (right column) are compared. Parameters in the fit may be found in Table \ref{Tab:DataCompare}. \textbf{Top row:} Test positive. Even visually it is clear that the quadratic correction improves the fit. The three outliers (red crosses) in the EU2020-fit have been excluded from the fit, as the testing was very irregular at that time. \textbf{Bottom row:} Growth rates (\ref{Eq:r1}-\ref{Eq:r3}) as a function of time. Note the big change over time which emphasizes the importance of this analysis, and note the similarities to Figure 2A of \cite{Liu}.}
	\label{Fig:DataCompare}
\end{figure*}

This invites the following interpretation. If we assume that the exponential part corresponds to the behavior exhibited by a normal SIR-model, then matrix $\beta A$ provides the spatial behavior. This is supported by the $\mu$-term providing non-conserved probability at each site, and $\beta A$ providing spreading through the network of sites. As an epidemic spreads, the mixing becomes greater, and it should therefore be assumed that $\frac{(\rho !)^2}{(2\rho)!}||A||^{\rho, a}$ decreases with $\rho$ for realistic systems, so the spatial function is a sub-exponential correction to the exponential growth. (Note that the spatial function is even and therefore cannot give pure exponential growth.)
With this in mind, I now propose the following:
\begin{theorem}\label{Conj:Afixed}
	The appropriate matrix $\beta A$ is fixed for a given societal structure. That is, two diseases that spread in the same society will have the same spatial function, provided that the method of infection is the same (e.g.\ airborne versus vector-borne) and that they target different age groups the same.
	This means that the coefficients of the spatial function from one disease may act as good prior for another.
\end{theorem}
To test both the model and the conjecture, I compare the functions
\begin{eqnarray}
	N_{\rm exp}(t) &=& N_0 e^{rt}\ ,\label{Eq:ThreeModels1}\\
	N_{\rm lin}(t) &=& N_0 e^{rt}\left(1 + c (t-t_0)\right)\ ,\label{Eq:ThreeModels2}\\
	N_{\rm qua}(t) &=& N_0 e^{rt}\left(1 + c (t-t_0)^2\right)\label{Eq:ThreeModels3}
\end{eqnarray}
to two points in the Danish part of the pandemic: The initial introduction of the EU2020 strains in March 2020 and the emergence of the omicron variant in late November of 2021.
Equation \eqref{Eq:ThreeModels1} represents a typical SIR-type model, Equation \eqref{Eq:ThreeModels3} represents a simplified version of the result in Equation \eqref{Eq:NFinal} with only 1 extra term, and Equation \eqref{Eq:ThreeModels2} represents an alternative model, where the correction is linear rather than quadratic. The quadratic correction also agrees with the branched polymers interpretation as it coincides with the intrinsic Hausdorff dimension \cite{Ambjoern}. (In other words, how much of the network is covered?)
Our goal, the growth rate, may be defined as
\begin{eqnarray}
	r(t) &=& \frac{d\ln N(t)}{dt}\ .
\end{eqnarray}
Applying this to the models (\ref{Eq:ThreeModels1}-\ref{Eq:ThreeModels3}) leads to
\begin{eqnarray}
	r_{\rm exp}(t) &=& r\ ,\label{Eq:r1}\\
	r_{\rm lin}(t) &=& r + \frac{c}{1 + c (t-t_0)}\ ,\label{Eq:r2}\\
	r_{\rm qua}(t) &=& r + \frac{ct/2}{1 + c (t-t_0)^2}\label{Eq:r3}\ .
\end{eqnarray}
Note the significance of this simple consideration. The growth rate may change over time because of the corrections, and, as can be seen in Figure \ref{Fig:DataCompare}, these changes may greatly impact the early estimate of transmission rate, which can make us misjudge the severity of a new virus or variant. Note also how the quadratic correction agrees qualitatively with Figure 2A of \cite{Liu}.

The exact choice of interval in the comparison of course impacts the fits. I have chosen 26th Feb - 31st Mar 2020 for EU2020 and 21st Nov - 6th Dec 2021 for omicron, as these are comparable in terms of infection numbers. This makes the 2021 interval rather small, but as omicron's growth rate is very large, it quickly mixes, so the correction terms are mostly only visible on this scale.

Conjecture \ref{Conj:Afixed} implies that the constant $c$ should be the same for every variant if the society remains constant. Unfortunately, lockdowns and restrictions make direct comparison between March 2020 and December 2021 impossible, but with access to the activity matrices \cite{EkspertgruppenMatricer} used for modeling the spread of covid-19 in Denmark \cite{EkspertgruppenRapporter}, an approximation may be made. Assuming a matrix of the form
\begin{eqnarray}\label{Eq:Aspace}
	A &=& A_{\rm NN} \otimes A_{\rm Act}
\end{eqnarray}
where $A_{\rm NN}$ is nearest neighbor interaction (1 if two sites are neighbors and 0 otherwise) and $A_{\rm Act}$ are the lockdown-dependent activity matrices, the norms may still be compared. The normalization of the matrices is chosen such that the largest eigenvalue is 1, i.e., with the growth rate scaled out (so only structure remains), and a $5\times 5$ grid is used for the spatial part. (If the site $a$ is chosen in the middle, the relevant coefficient $||A||^{\rho=1, a}$ does not change with increased grid size beyond $5\times 5$, so this is sufficient.)

Because the model still has a free parameter $\beta$, the ratio of the coefficients is compared. Using 10-year age groups, the norm ratio is in agreement with the coefficients of the quadratic corrections
\begin{align}
	\begin{split}
		\frac{||A_{\rm Om}||_{\rm fer}^{\rho = 1, a}}{||A_{\rm EU20}||_{\rm fer}^{\rho = 1, a}}\ &=\ 0.77 \pm 0.12\\
		\frac{\left(r_{\rm Om} / (e^{r_{\rm Om}/2}-1)\right)^2 c_{\rm Om}}{\left(r_{\rm EU20} / (e^{r_{\rm EU20}/2}-1)\right)^2 c_{\rm EU20}}\ &=\ 0.83
	\end{split}
\end{align}
which supports Conjecture \ref{Conj:Afixed}. The coefficient depending on $r$ comes from $\mu$ in Equation \eqref{Eq:NFinal}. An interval is given for the norm ratio, as it varies slightly by which age group the index $a$ corresponds to. (Only $a$ corresponding to groups 20-69 is used for this interval, as they are the main actors in these time periods.) A fermionic norm is used, because this displays the most realistic spatial behavior, see the supplementary material for details.

\begin{table*}[!htbp]
	\centering
	\caption{Comparison of likelihood fits to the models (\ref{Eq:ThreeModels1}-\ref{Eq:ThreeModels3}). To account for over-dispersion, a negative binomial distribution is used for the likelihood optimization, where the size-parameter is also optimized. Using AIC-rule that an improvement of 1 justifies another parameter \cite{nll}, the quadratic correction constitutes a significant improvement on the exponential fit. The norms from the model \eqref{Eq:NFinal} using \eqref{Eq:Aspace} are also given. A range is given for the norms as they depend on the age group that the index $a$ corresponds to (here the age range 20-69 is used). The ratios between the coefficients of the quadratic corrections and the norms are in agreement which supports Conjecture \ref{Conj:Afixed}.}
	\begin{tabular}{*7c}
		\toprule
		\textbf{Variant} &  \multicolumn{3}{c}{\textbf{EU2020}} & \multicolumn{3}{c}{\textbf{Omicron}}\\
		\midrule[0.7pt]
		\textbf{Correction to Exp} & \textbf{None} & \textbf{Linear} & \textbf{Quadratic} & \textbf{None} & \textbf{Linear} & \textbf{Quadratic}\\
		\midrule
		\textbf{Negative log-likelihood} & 134.0 & 127.7 & 124.3 & 59.7 & 60.1 & 52.8\\
		\midrule[0.1pt]
		\textbf{Growth rate $r=2\re(\mu)$} & 0.22 & 0.13 & 0.0067 & 0.36 & 0.31 & 0.17 \\
		\midrule[0.1pt]
		\textbf{Coefficient $c$} & {} & 0.25   & 0.20  & {} & 6.2 & 0.18\\
		\midrule[0.7pt]
		\textbf{$||A||_{\rm fer}^{\rho = 1, a}$} &  \multicolumn{3}{c}{$0.058\pm 0.025$}  & \multicolumn{3}{c}{$0.049\pm 0.023$} \\
		\midrule[0.7pt]
		\textbf{$\frac{\left(\frac{e^{\re(\mu)} - 1}{\re(\mu)}\right)^2||A||_{\rm fer}^{\rho = 1, a}}{c_{\rm qua}}$} &  \multicolumn{3}{c}{$0.29\pm 0.13$}  & \multicolumn{3}{c}{$0.28\pm 0.13$} \\
		\bottomrule
	\end{tabular}
	\label{Tab:DataCompare}
\end{table*}

\section{Conclusion and Discussion}
In this paper I presented a spatial epidemic model in a second quantization framework in order to help establish priors for early estimates of the infectiousness of new diseases and variants. It results in a spatial correction to the traditional exponential growth, which is shown to agree with data from the covid-19 epidemic in Denmark and a previous simulation study.

This second quantization framework is enticing, because a connection to solid state physics unlocks tools from a completely different field and hopefully encourages scientists from solid state physics to apply their methods in epidemic modeling.

While the quadratic correction certainly provides a significant improvement on the fit, this should not be taken as final proof of viability of the model in Equation \eqref{Eq:Hamiltonians}. That it coincides with known results from branched polymers is striking, but a quadratic correction may a priori come from other effects too, and more support is therefore needed. It does, however, highlight the importance of spatial effects when estimating the infectiousness of a new disease or variant. Under no circumstances should an exponential fit be used blindly.

\rule{\textwidth/3}{1pt}
\acknowledgements
\paragraph{Acknowledgements.} Data and activity matrices were kindly provided by Statens Serum Institut. I would also like to thank L.\ E.\ Christiansen, K.\ Splittorff, and M.\ H.\ Appel for interesting discussions.

\end{document}


\newcommand{\be}{\begin{eqnarray}}
		\newcommand{\ee}{\end{eqnarray}}
	\newcommand{\del}{\partial}
	\newcommand{\nn}{\nonumber}
	\newcommand{\STr}{{\rm Str}}
	\newcommand{\Sdet}{{\rm Sdet}}
	\newcommand{\Pf}{{\rm Pf}}
	\newcommand{\mat}{\left ( \begin{array}{cc}}
		\newcommand{\emat}{\end{array} \right )}
	\newcommand{\vect}{\left ( \begin{array}{c}}
		\newcommand{\evect}{\end{array} \right )}
	\newcommand{\tr}{{\rm Tr}}
	\newcommand{\hm}{\hat m}
	\newcommand{\ha}{\hat a}
	\newcommand{\hz}{\hat z}
	\newcommand{\hze}{\hat \zeta}
	\newcommand{\hx}{\hat x}
	\newcommand{\hy}{\hat y}
	\newcommand{\tm}{\tilde{m}}
	\newcommand{\ta}{\tilde{a}}
	\newcommand{\U}{\rm U}
	\newcommand{\D}{\slashed{D}}
	\newcommand{\hc}{^\dagger}
	\newcommand{\inv}{^{-1}}
	\newcommand{\diag}{{\rm diag}}
	\newcommand{\sign}{{\rm sign}}
	\newcommand{\ct}{\tilde{c}}

\title{
{\it  Supplementary Material to:}\\
On the Role of Spatial Effects in Early Estimates of Disease Infectiousness:\\A Second Quantization Approach}

\author{Adam Mielke}\email{admi@dtu.dk}\affiliation{Dynamical Systems, Technical University of Denmark, Asmussens Allé, 303B, 2800 Kgs.\ Lyngby, Denmark}
 
\date{\today}
\maketitle

\begin{appendix}
\section{Detailed Solution of Model}
\subsection{Derivation of the Particle Number Time Evolution}
Starting with 
\begin{eqnarray}\label{Eq:Hamiltonians}
	H_{\rm fer} &=& \beta \sum_{jk} c\hc_j A_{jk} N_k + \gamma\sum_j c_j + \frac{1}{2}\left(\mu\sum_{jk} c\hc_j B_{jk} c_k + h.c.\right)\ ,\\
	H_{\rm bos} &=& \beta \sum_{jk} b\hc_j A_{jk} N_k + \gamma\sum_j N_j + \frac{1}{2}\left(\mu\sum_{jk} b\hc_j B_{jk} b_k + h.c.\right)\ .
\end{eqnarray}
and the relations
\begin{eqnarray}
	\left\{c_j,\ c_k\right\}\ =\ \left\{c\hc_j,\ c\hc_k\right\}\ =\ 0\ &,&\	\left\{c_j,\ c\hc_k\right\}\ =\ \delta_{jk}\\
	\left[b_j,\ b_k\right]\ =\ \left[b\hc_j,\ b\hc_k\right]\ =\ 0\ &,&\ \left[b_j,\ b\hc_k\right]\ =\ \delta_{jk}\\
	\left[N_j,\ c_k\right] = -\delta_{jk}c_j\ &,&\ \left[N_j,\ c\hc_k\right] = \delta_{jk}c\hc_j\label{Eq:Fer:Com}\\
	\left[N_j,\ b_k\right] = -\delta_{jk}b_j\ &,&\ \left[N_j,\ b\hc_k\right] = \delta_{jk}b\hc_j\label{Eq:Bos:Com}\ ,
\end{eqnarray}
I will derive the time evolution
\begin{eqnarray}
	N(\tau) &=& \sum_m \bra{\psi(0)} e^{H\hc \tau^*} c\hc_m c_m e^{H\tau} \ket{\psi(0)}\ .
\end{eqnarray}
Up until the derivation of the norm, I will use the fermionic model as an example, but the bosonic version follows the same way because of the similar structure of \eqref{Eq:Fer:Com} and \eqref{Eq:Bos:Com}. I therefore also drop the subscript on $H_{\rm fer}$ for now.

\paragraph*{Simplified Case With Only Transport}
Let me start with $\beta = \gamma = 0$ as a warm-up
\begin{eqnarray}
	H &=& \sum_{jk}\left(\mu B_{jk} + \mu^* (B_{jk})\hc\right)c_j\hc c_k\ .
\end{eqnarray}
Here the commutator is trivial
\begin{eqnarray}
	\sum_m \left[N_m,\ H\right] &=& \sum_{jkm}\left(\mu B_{jk} + \mu^* (B_{jk})\hc\right) \left[c\hc_m c_m,\ c_j\hc c_k\right]\nn\\
	&=& \sum_{jkm}\left(\mu B_{jk} + \mu^* (B_{jk})\hc\right)  \left(c_j\hc \left[c\hc_m c_m,\ c_k\right] + \left[c\hc_m c_m,\ c_j\hc\right]  c_k \right) \nn\\
	&=& \sum_{jkm}\left(\mu B_{jk} + \mu^* (B_{jk})\hc\right)  \left(- \delta_{mk} c_j\hc c_k + \delta_{mj} c_j\hc c_k \right) \nn\\
	&=& \sum_{jkm}\left( -c_j\hc c_m\left(\mu B_{jm} + \mu^* (B_{jm})\hc\right) + c_m\hc c_k\left(\mu B_{mk} + \mu^* (B_{mk})\hc\right) \right) \nn\\
	&=& 0\ ,
\end{eqnarray}
and the time evolution of the counting operator becomes
\begin{eqnarray}
	N(\tau) &=& N(0) \bra{\psi(0)}\sum_{\rho=0}^{\infty}\frac{1}{\rho!}\left(\left(\tau+\tau^*\right)\sum_{jk}\left(\mu B_{jk} + \mu^* (B_{jk})\hc\right)c_j\hc c_k\right)^{\rho}\ket{\psi(0)} \ .
\end{eqnarray}
The surviving terms are the ones that have adjacent creation and annihilation operators. (We can only move a particle from sites that have a particle.) In other words, the following is obtained
\begin{eqnarray}
	N(\tau) &=& N(0) \mathbf{1}_a e^{2\re(\tau) \left(\mu B + \mu^* B\hc\right) } \mathbf{1}_a^T\ ,
\end{eqnarray}
where $\mathbf{1}_a$ is a row vector that is 0 everywhere but $a$ and the exponential is in the matrix sense.
This is where the treatment of imaginary time becomes very important. In the Euclidean theory, where $\tau\in\mathbf{R}$, there is an influx/loss of particles, whereas the particle number remains constant for $i\tau = t\in\mathbf{R}$.\\

\paragraph*{Simplified Case with Only Infection.}
Because the norms are the most complicated part of the derivation, I will start by showing them in the simplest case where they appear: $\mu = \gamma = 0$.
Here the commutator is
\begin{eqnarray}
	\sum_m\left[N_m, H\right] &=&  \beta\sum_{jk} A_{kj} c\hc_k N_j = H\ ,
\end{eqnarray}
which follows directly from \eqref{Eq:Fer:Com}.
The counting operator can now be moved through the exponential. In each term of the expansion, a factor equal to the power is acquired, which leaves
\begin{eqnarray}
	N(\tau) &=& \bra{\psi(0)} \sum_m e^{H\hc \tau^*} e^{H \tau} N_m  + e^{H\hc \tau^*} e^{H\tau} H\tau\ket{\psi(0)}\nn\\
	&=& \bra{\psi(0)}  e^{H\hc \tau^*} e^{H \tau} N(0)  + e^{H\hc \tau^*} e^{H\tau} H\tau\ket{\psi(0)}\ .
\end{eqnarray}
Keep in mind that $[H\hc,\ H] \neq 0$, so the exponentials are not combined. The expansion is written out
\begin{eqnarray}
	N(\tau) &=& N(0)\bra{\psi(0)} \left(\sum_{\rho=0}^{\infty} \frac{1}{\rho!}\left(H\hc \tau^*\right)^\rho\right) \left(\sum_{\rho=0}^{\infty} \frac{1}{\rho!}\left(H \tau\right)^\rho\right) \ket{\psi(0)}\nn\\
	&&+ \bra{\psi(0)}\left(\sum_{\rho=0}^{\infty} \frac{1}{\rho!}\left(H\hc \tau^*\right)^\rho\right) \left(\sum_{\rho=0}^{\infty} \frac{1}{\rho!}\left(H \tau\right)^{\rho+1}\right) \ket{\psi(0)}\ .
\end{eqnarray}
Note that only $H$ creates additional particles and only $H\hc$ removes them. Since states with different particle numbers are orthogonal, only terms with an equal amount of $H$ and $H\hc$ will survive. This means the sums collapse to
\begin{eqnarray}
	N(\tau) &=& N(0)\bra{\psi(0)} \sum_{\rho=0}^{\infty} \frac{|\tau|^{2\rho}}{\left(\rho!\right)^2}{H\hc}^\rho {H}^\rho\ket{\psi(0)}\nn\\
	&&+ \bra{\psi(0)}\sum_{\rho=0}^{\infty} \frac{|\tau|^{2(\rho+1)}}{(\rho+1)! \rho!}{H\hc}^{\rho+1} {H}^{\rho+1} \ket{\psi(0)}\ ,
\end{eqnarray}
and the problem is reduced to expectation values of the form
\begin{eqnarray}
	\bra{\psi(0)} {H\hc}^\rho {H}^\rho\ket{\psi(0)} &=& \bra{\psi(0)} \left(\beta^*\sum_{jk} A^*_{jk} N_j c_k\right)^\rho \left(\beta\sum_{jk} A_{jk} c\hc_j N_k\right)^\rho\ket{\psi(0)}
\end{eqnarray}
Applying the initial condition
\begin{eqnarray}\label{Eq:InitCond}
	\ket{\psi(0)} = c\hc_a\ket{0}\ ,
\end{eqnarray}
the matrix norm appears as
\begin{eqnarray}
	||A||_{\rm fer}^{\rho, a} &=& \bra{0}c_a \left(\sum_{jk} A^*_{jk} N_j c_k\right)^\rho \left(\sum_{jk} A_{jk} c\hc_j N_k\right)^\rho c\hc_a\ket{0}
\end{eqnarray}
The general conditions for non-zero terms in the both brackets are
\begin{eqnarray}
	\sum_{\substack{p_j = 1...n\\ p_{2k-1} = p_1...p_{2k-2}\\p_{2k} \neq p_1...p_{2k-1}} }  A_{p_{2\rho}p_{2\rho-1}} ... A_{p_{2}p_{1}}
\end{eqnarray}
where $p_1 = a$. In other words, the odd indices $p_{2k-1}$ must equal an index that appears somewhere before in the chain, or the counting operator will make the term 0. Similarly, the even indices $p_{2k}$ cannot appear before, or the duplicate creation operators will cancel each other.

The two brackets are of course the same up to Hermitian conjugate. The matching terms between the brackets are all $p_{2k}' \in \sigma(p_{2k})$, where $\sigma$ denotes the permutation group. That is, all terms with the same creation/annihilation operators. There is additionally a sign from the permutation of these operators because of the anti-commutation. So the full time evolution for the initial condition \eqref{Eq:InitCond} is
\begin{eqnarray}\label{Eq:FermionFinal}
	N(\tau) &=& N(0)\sum_{\rho=0}^{\infty} \frac{(|\beta| |\tau|)^{2\rho}}{\left(\rho!\right)^2} ||A||_{\rm fer}^{\rho, a} + \sum_{\rho=0}^{\infty} \frac{(|\beta| |\tau|)^{2(\rho+1)}}{(\rho+1)! \rho!} ||A||_{\rm fer}^{\rho+1, a}\nn\\
	&=& \sum_{\rho=0}^{\infty} \frac{\left(N(0) + \rho\right)(|\beta| |\tau|)^{2\rho}}{(\rho!)^2} ||A||_{\rm fer}^{\rho, a}\ ,
\end{eqnarray}
where the norm is
\begin{eqnarray}\label{Eq:FermionNorm}
	||A||_{\rm fer}^{\rho, a} &=& \sum_{\substack{p_j, q_j = 1...n\\ p_{2k-1} = p_1...p_{2k-2}\\p_{2k} \neq p_1...p_{2k-1}\\ q_{2k-1} = q_1...q_{2k-2}\\ q_{2k} \in \sigma(p_{2k}) }} \sign(\sigma) A^*_{q_{2\rho}q_{2\rho-1}} ... A^*_{q_{2}q_{1}} A_{p_{2\rho}p_{2\rho-1}} ... A_{p_{2}p_{1}}
\end{eqnarray}
and
\begin{eqnarray}
	||A||_{\rm fer}^{\rho=0, a} &=& 1
\end{eqnarray}
by definition. $||A||_{\rm fer}^{\rho, a}$ is real by construction for all values of $A$, $\rho$, and $a$.
Note here that all terms appear, regardless of whether $\tau$ or $t$ is real. As this term inherently does not conserve particle number, this is very natural.

In some cases, the following form may also be of use. View
\begin{eqnarray}
	C_{jk} &=& c\hc_j c_j c\hc_k
\end{eqnarray}
as a matrix and note that
\begin{eqnarray}
	C^2 &=& 0\ .
\end{eqnarray}
The problem may thus be simplified if the commutator $\left[C,\ A\right]$ is known. Writing the norm as a matrix product
\begin{eqnarray}\label{Eq:AC_commutator}
	||A||_{\rm fer}^{\rho, a} &=& \bra{0} \mathbf{1}_a \left(C\hc A\hc\right)^{\rho} \mathbf{c}^T \mathbf{c\hc} \left(A C\right)^{\rho} \mathbf{1}_a^T\ket{0}
\end{eqnarray}
where $\mathbf{c}$ and $\mathbf{c\hc}$ are row vectors containing the corresponding operators, and the transposition works in the same space as $A$. (This means that $ \mathbf{c}^T \mathbf{c\hc} $ is a matrix.) $\mathbf{1}_a$ is a row vector that is 0 everywhere but $a$.
The chains of matrices may be reduced to
\begin{eqnarray}\label{Eq:AC_commutator2}
	||A||_{\rm fer}^{\rho, a} &=& \bra{0} \mathbf{1}_a C\hc \left[A\hc,\ C\hc\right]^{\rho-1} A\hc \mathbf{c}^T \mathbf{c\hc} A \left[C,\ A\right]^{\rho-1} C \mathbf{1}_a^T\ket{0}\ .
\end{eqnarray}\\

\paragraph*{Bosonic Norm.}
Before moving on to the case considered in the main text, the bosonic case must briefly be mentioned as well. Note that there is no sign changes, because the operators commute rather than anti-commute, but we have to take into account the factors from the creation and annihilation operators, which are no longer only 0 or 1. We may express it formally the following way
\begin{eqnarray}\label{Eq:BosonNorm}
	||A||_{\rm bos}^{\rho, a, \nu} &=& \sum_{\substack{p_j, q_j = 1...n\\ q_{2k} \in \sigma(p_{2k}) }} \left(A_{q_{2\rho}q_{2\rho-1}} \xi_{q_{2\rho}q_{2\rho-1}}\right) ... \left(A_{q_{2}q_{1}}^* \xi_{q_{2}q_{1}}\right)  \left(A_{p_{2\rho}p_{2\rho-1}}\xi_{p_{2\rho}p_{2\rho-1}} \right) ... \left(A_{p_{2}p_{1}} \xi_{p_{2}p_{1}}\right)\nn\\
\end{eqnarray}
The parentheses are only there to guide the eye. The factors $\xi$ are defined recursively
\begin{eqnarray}\label{Eq:BosonFactors}
	\xi_{p_{2j}p_{2j-1}} &=& \begin{cases}
		\sqrt{\sum_{m = 0}^{j-1} \delta_{p_{2j}, p_{2m}}} \left(\sum_{m = 1}^{j-1} \delta_{p_{2j-1}, p_{2m}}\right)  \ &,\ \sum_{m = 0}^{j-1} \delta_{p_{2j}, p_{2m}} < \nu_j \\
		0\ &,\ {\rm otherwise}
	\end{cases}\ .
\end{eqnarray}
For the sake of the factors, I define $p_0 = a$. Equation \eqref{Eq:BosonNorm} is of course slight abuse of notation as each element of $\xi$ depends on the configuration of the sums in Equation \eqref{Eq:BosonNorm}, but it is kept in this form for readability.
Note that the one-level model, i.e. $\nu_j=1$ for all $j$, coincides with the fermionic model apart from the sign of the permutation. However, these signs play an important part, see the Section \ref{Sec:FerBosDiff}.
\\

\paragraph*{Infection and Chemical Potential.}
This is the case $\gamma = 0$ and $B_{jk} = \delta_{jk}$, which considered in the main text. That is,
\begin{eqnarray}
	H &=& \beta \sum_{jk} c\hc_k A_{kj} N_j + \re(\mu)\sum_{j}c_j\hc c_j\nn\\
	&\equiv& H_0 + \re(\mu)\sum_j N_j \ .
\end{eqnarray}
As mentioned there, the trick is that the commutator is recursive
\begin{eqnarray}
	\left[\re(\mu)\widehat{N}, H_0\right] &=& \re(\mu)H_0\ .
\end{eqnarray}
I adapt a special case of the Baker-Campbell-Hausdorff formula \cite{HallHandbook}
\begin{eqnarray}
	e^X e^Y &=& e^{X + \frac{s}{1-e^{-s}}Y}\ {\rm for}\ \left[X,\ Y\right] = sY
\end{eqnarray}
in the following way. Note that 
\begin{eqnarray}
	e^{-X}e^{-\exp(s)Y} &=& e^{-X + \frac{s e^{s}}{1-e^{s}}Y}\ ,
\end{eqnarray}
which implies
\begin{eqnarray}
	e^{X}e^{Y} &=& e^{\exp(s) Y} e^{X}\ {\rm for}\ \left[X,\ Y\right] = sY\ .
\end{eqnarray}
I make another trivial rescaling
\begin{eqnarray}
	e^{Y + X}\ =\ e^{X + Y}\ =\ e^X e^{\frac{1-e^{-s}}{s}Y}\ =\ e^{\frac{e^{s}-1}{s}Y} e^X\ {\rm for}\ \left[X,\ Y\right] = sY
\end{eqnarray}
which in this case becomes
\begin{eqnarray}
	e^{\tau (H_0 + \re(\mu)\widehat{N})} &=& e^{\tau\frac{e^{\re(\mu)} - 1}{\re(\mu)}H_0} e^{\tau\re(\mu)\sum_j N_j}\ .
\end{eqnarray}
As also noted in the main text, this allows the counting operator to be applied independently of $H_0$, which leaves
\begin{eqnarray}\label{Eq:FermionFinalmu}
	N(\tau) &=& e^{2\re(\mu)\re(\tau)N(0)}\sum_m \bra{\psi(0)} e^{\frac{e^{\re(\mu)} - 1}{\re(\mu)} H_0\hc \tau^*} N_m e^{\frac{e^{\re(\mu)} - 1}{\re(\mu)}H_0\tau} \ket{\psi(0)}\nn\\
	&=& e^{2\re(\mu)\re(\tau)N(0)} \sum_{\rho=0}^{\infty} \frac{\left(N(0) + \rho\right)\left(\frac{e^{\re(\mu)} - 1}{\re(\mu)}|\beta| |\tau|\right)^{2\rho}}{(\rho!)^2} ||A||^{\rho, a}\ .
\end{eqnarray}
For $\mu = 0$, this reduces to \eqref{Eq:FermionFinal}. The quantum interpretation of the result \eqref{Eq:FermionFinalmu} seems the following. The number of occupied states is not really influenced by $\mu$. Instead the increase in particle number comes from the non-conserved probability of the non-Hermitian Hamiltonian. So if $\mu \sum_j N_j$ provides the pure exponential growth at a point, then the corrections from spatial behavior is given by an even function with coefficients depending on $\beta A$.

\subsection{Difference Between Fermionic and One-level Bosonic Models}\label{Sec:FerBosDiff}
\begin{figure}
	\centering
	\includegraphics[width=0.25\textwidth]{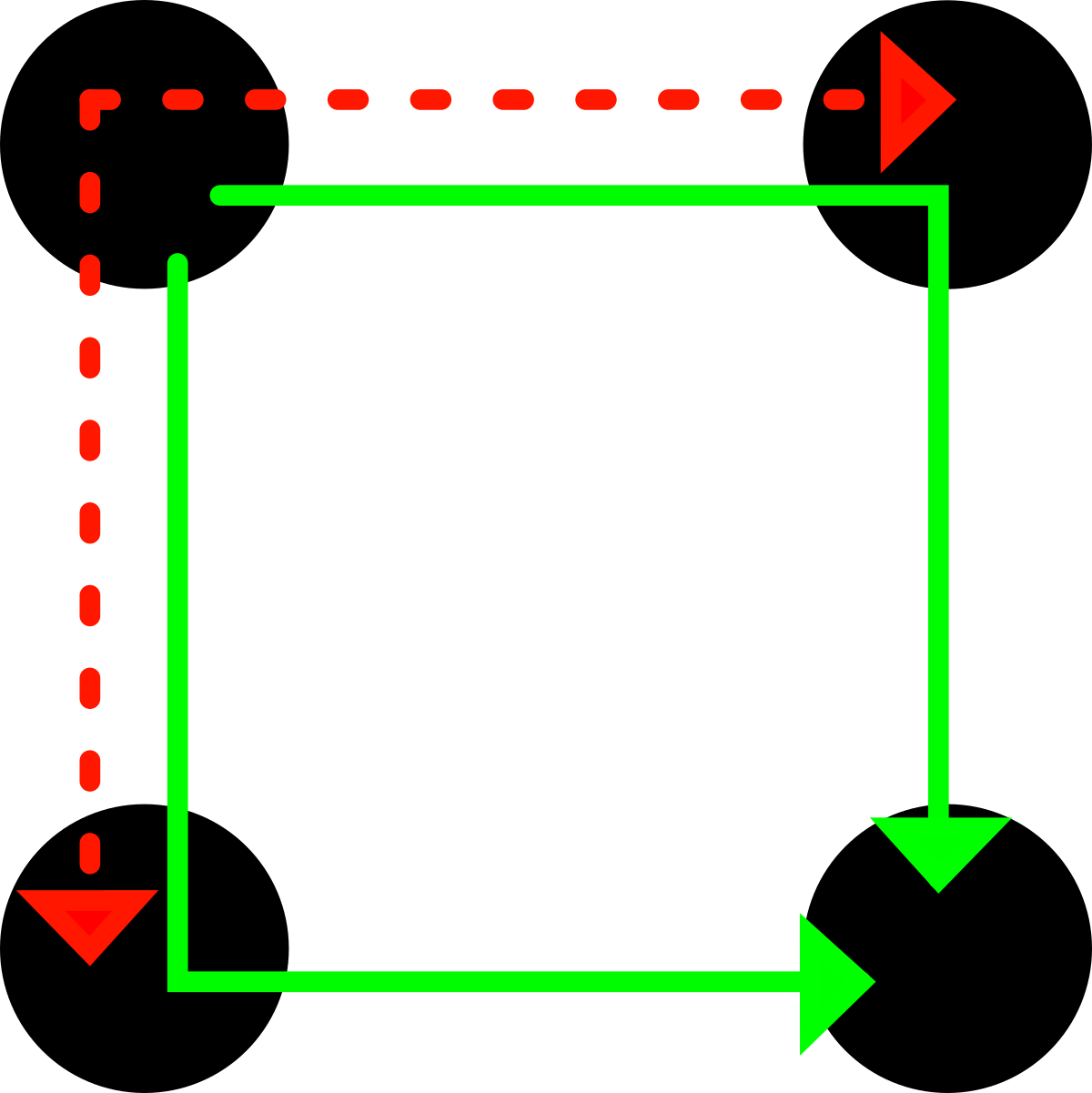}
	\caption{Graphic representation of the difference between the fermionic and one-level bosonic models. Depicted is the contributions to $\rho=2$ for the system described in Section \ref{Sec:FerBosDiff}. The one-level bosonic model has all terms, whereas parity excludes the dashed red line in the fermionic model. This difference in the cross-term is why I claim that the Fermionic model is more physical when considering individuals.}
	\label{Fig:BosFer}
\end{figure}

To illustrate the difference between the fermionic and one-level bosonic models, which also hints at the general difference between fermioninc and bosonic models, let us consider a very basic finite-$n$ configuration: 4 sites arranged in a square with nearest neighbor interaction
\begin{eqnarray}
	A &=& \left(\begin{matrix}
		0 & A_{12} & A_{13} & 0\\
		A_{21} & 0 & 0 & A_{24}\\
		A_{31} & 0 & 0 & A_{34}\\
		0 & A_{42} & A_{43} & 0
	\end{matrix}\right)
\end{eqnarray}
and start at $a=1$.
Here we have
\begin{align}
	\begin{split}
		||A||_{\rm bos}^{\rho = 1, a=1, \nu = 1} = ||A||_{\rm fer}^{\rho = 1, a=1} &= |A_{21}|^2 + |A_{31}|^2\\
		||A||_{\rm fer}^{\rho = 2, a=1} &= |A_{42}|^2 |A_{21}|^2 + |A_{43}|^2 |A_{31}|^2\\
		||A||_{\rm bos}^{\rho = 2, a=1, \nu = 1} &= |A_{42}|^2 |A_{21}|^2 + |A_{43}|^2 |A_{31}|^2 + 4|A_{31}|^2 |A_{21}|^2
	\end{split}\ .
\end{align}
So what is the term $4|A_{31}|^2 |A_{21}|^2$ that is left out in the Fermionic model? It is the movement to both nodes at the same time. That is, there is no cross-term between $1\to 2$ and $1\to 3$, which is what makes the fermionic model preferable for modeling of individuals. See Figure \ref{Fig:BosFer} for a graphical representation of this.

\section{Special Cases of Matrix Models}
I would like to provide some examples of null-models that may be useful.
\subsection{Complete Mixing}
If we consider full mixing to be
\begin{eqnarray}
	A_{jk} &=& \frac{1}{\sqrt{n-1}}\ ,\ j\neq k\ ,
\end{eqnarray}
where the factor $1/\sqrt{n-1}$ is there to make the scaling of $||A||^{1, a}$ natural, the fermionic norm reduces to the sum over the sign of the permutation group, which is 0 for more than one element. This means that only $\rho = 0,1$ survives, and we are left with
\begin{eqnarray}
	N_{\rm full\,mix,\,fer}(\tau) &=& e^{2\re(\mu)\re(\tau)N(0)}\left(1 + 2\left(\frac{e^{\re(\mu)} - 1}{\re(\mu)}|\beta| |\tau|\right)^{2} ||A||_{\rm fer}^{1, a}\right)\nn\\
	&=& e^{2\re(\mu)\re(\tau)}\left(1 + 2 \left(\frac{e^{\re(\mu)} - 1}{\re(\mu)}|\beta| |\tau|\right)^{2} \right)\ .
\end{eqnarray}
The one-level bosonic version may also be solved explicitly.
The number of particles that can be chosen is
\begin{eqnarray}
	\frac{(n-1)!}{(n-1-\rho)!}\ .
\end{eqnarray}
Keep in mind that I do not divide by $\rho!$, because the order matters here. The $-1$ comes from the fixed entry $a$. The counting operators can then be chosen in $\rho!$ ways, as it has to match the previous creation operators. There are $\frac{(n-1)!}{(n-1-\rho)!\rho!}$ unique particle combinations that may be matched. Because of symmetry, these appear equally frequently. This means there are
\begin{eqnarray}
	\frac{\frac{(n-1)!}{(n-1-\rho)!}}{\frac{(n-1)!}{(n-1-\rho)!\rho!}} &=& \left(\rho!\right)^2\ .
\end{eqnarray}
configurations per unique particle configuration, and thus $\left(\rho!\right)^4$ combinations of $H$ and $H\hc$ per particle configuration.
So in total there are
\begin{eqnarray}
	\frac{(n-1)!}{(n-1-\rho)!\rho!} (\rho!)^4 &=& \frac{(n-1)!}{(n-1-\rho)!} (\rho!)^3
\end{eqnarray}
terms, each equal to $\frac{1}{(n-1)^\rho}$. The norms very quickly become tricky to calculate numerically, and the unfavorable scaling can be seen here.
The final result is 
\begin{eqnarray}
	N_{\rm full\,mix,\,bos}(\tau) &=& e^{2\re(\mu)\re(\tau)} \sum_{\rho=0}^{\infty} \frac{\left(1 + \rho\right)\left(\frac{e^{\re(\mu)} - 1}{\re(\mu)}|\beta| |\tau|\right)^{2\rho}}{(n-1)^\rho} \frac{(n-1)!}{(n-1-\rho)!} \rho! \ .
\end{eqnarray}
Note that the limit $n\to\infty$ does not make sense here as it diverges in $\rho$. This can be seen using Stirling's approximation
\begin{eqnarray}
	n! &\overset{n\gg 1}{\propto}& e^{n\ln(n)-n}\ ,
\end{eqnarray}
and assuming that $n$ increases faster than any $\rho$, the large $n$-limit can be found
\begin{eqnarray}
	N_{\rm full\,mix,\,bos}(\tau) &\overset{n\gg 1}{=}& e^{2\re(\mu)\re(\tau)} \sum_{\rho=0}^{\infty} \frac{\left(1 + \rho\right)\left(\frac{e^{\re(\mu)} - 1}{\re(\mu)}|\beta| |\tau|\right)^{2\rho}}{(n-1)^\rho} \rho!\nn\\
	&&\times\exp\left\{(n-1)\ln(n-1) - (n-1-\rho)\ln(n-1-\rho) -\rho\right\}\nn\\
	&\overset{n\gg \rho}{=}& e^{2\re(\mu)\re(\tau)} \sum_{\rho=0}^{\infty} \left(1 + \rho\right)\left(\frac{e^{\re(\mu)} - 1}{\re(\mu)}|\beta| |\tau|\right)^{2\rho} \rho! e^\rho\ .
\end{eqnarray}
It is possible that there exists a critical rescaling of the matrix elements that allows a finite, non-zero limit, though this is not straightforward as $A$ cannot depend on $\rho$.

\subsection{Random Models}
\begin{figure}[t]
	\centering
	\includegraphics[width=0.3\textwidth]{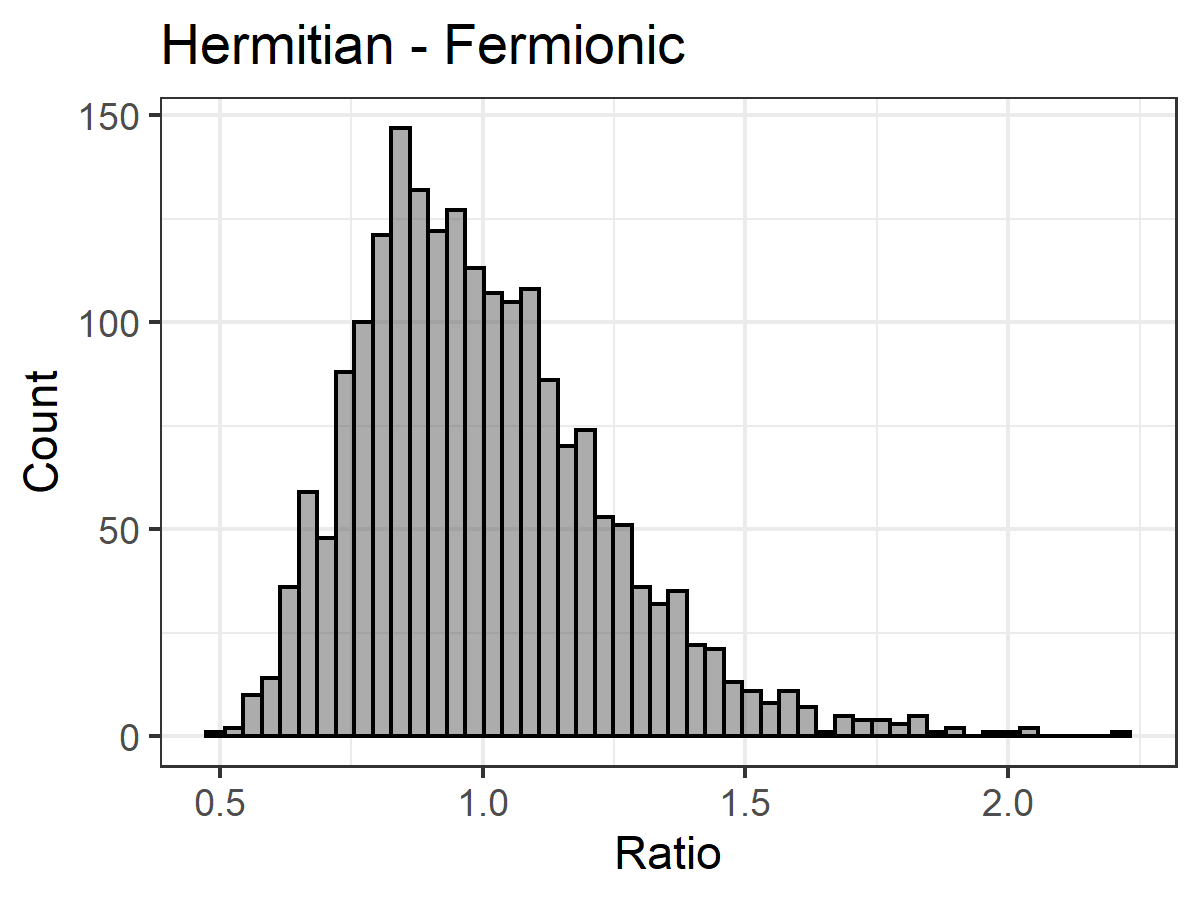}
	\includegraphics[width=0.3\textwidth]{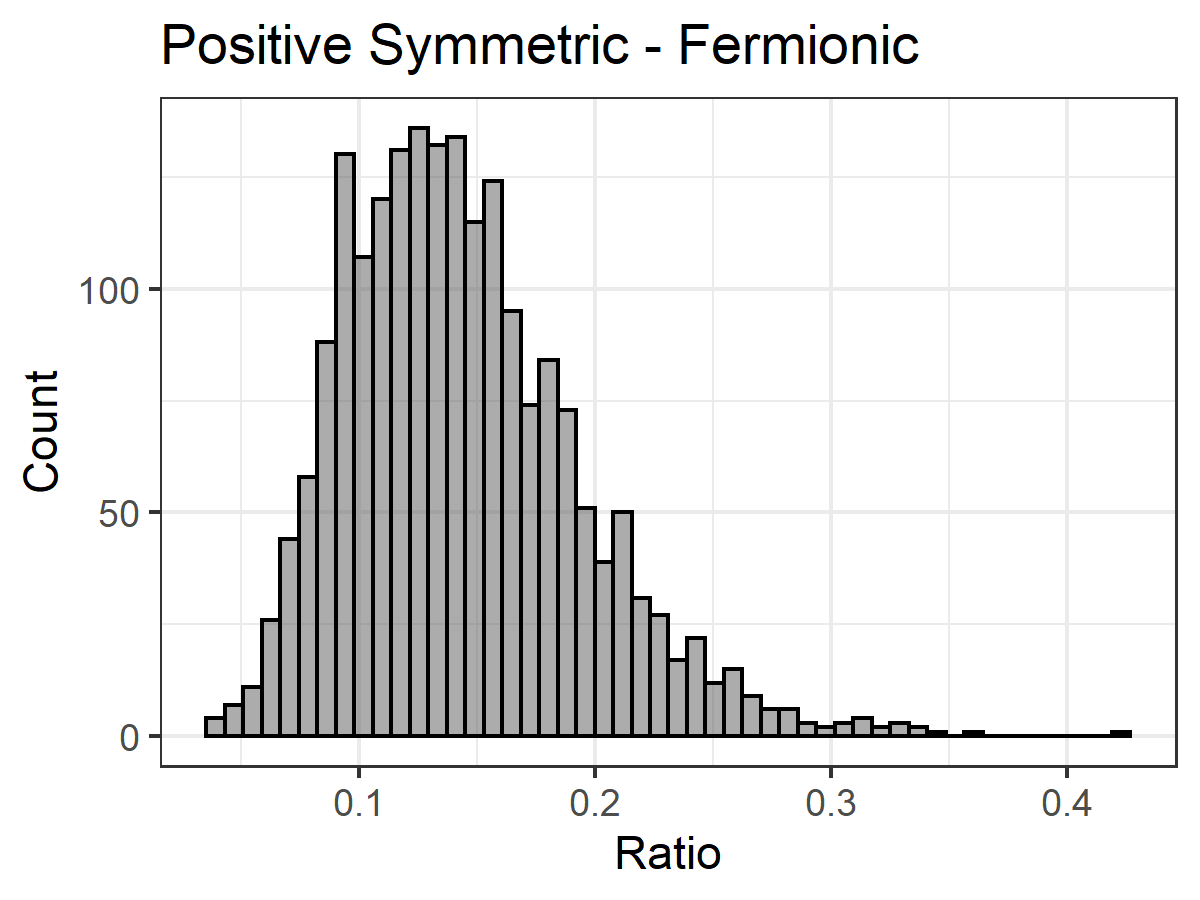}
	\includegraphics[width=0.3\textwidth]{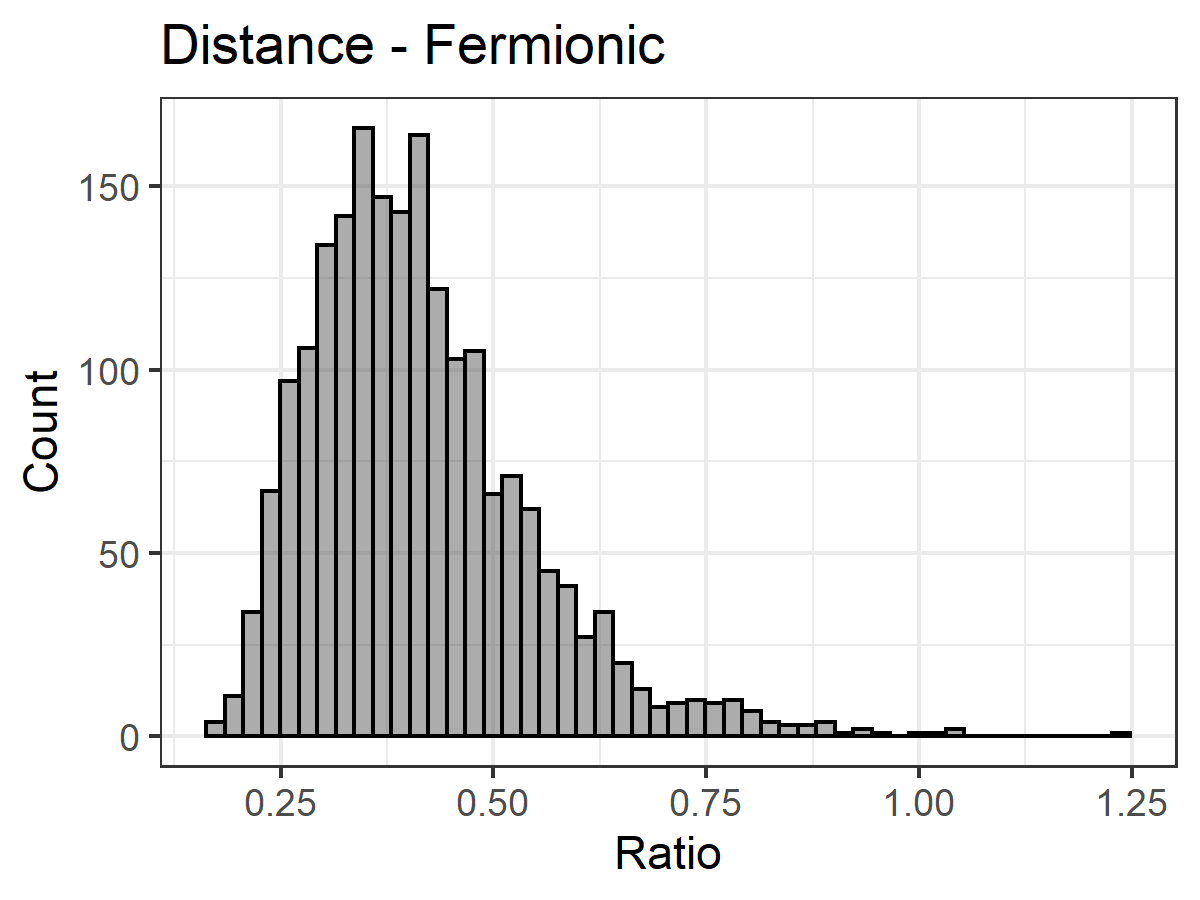}\\
	\includegraphics[width=0.3\textwidth]{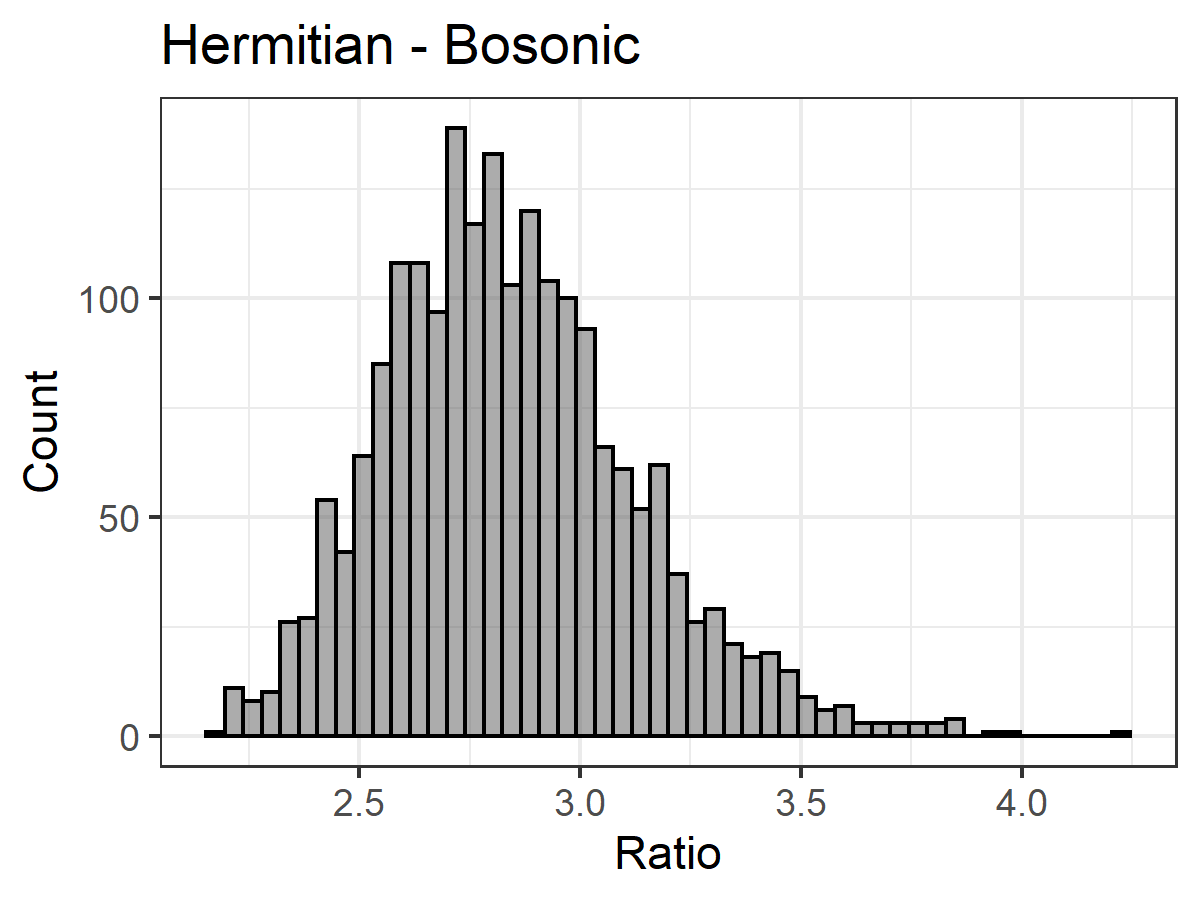}
	\includegraphics[width=0.3\textwidth]{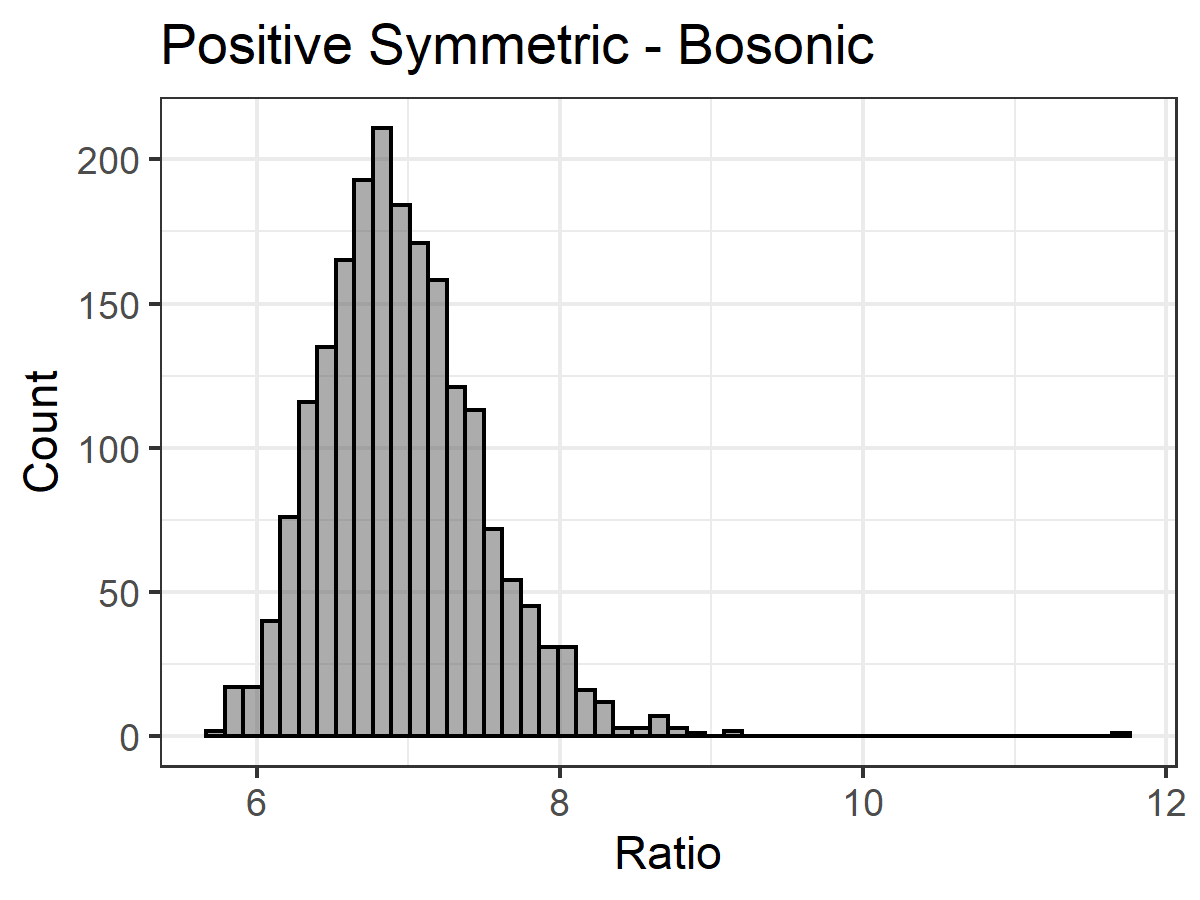}
	\includegraphics[width=0.3\textwidth]{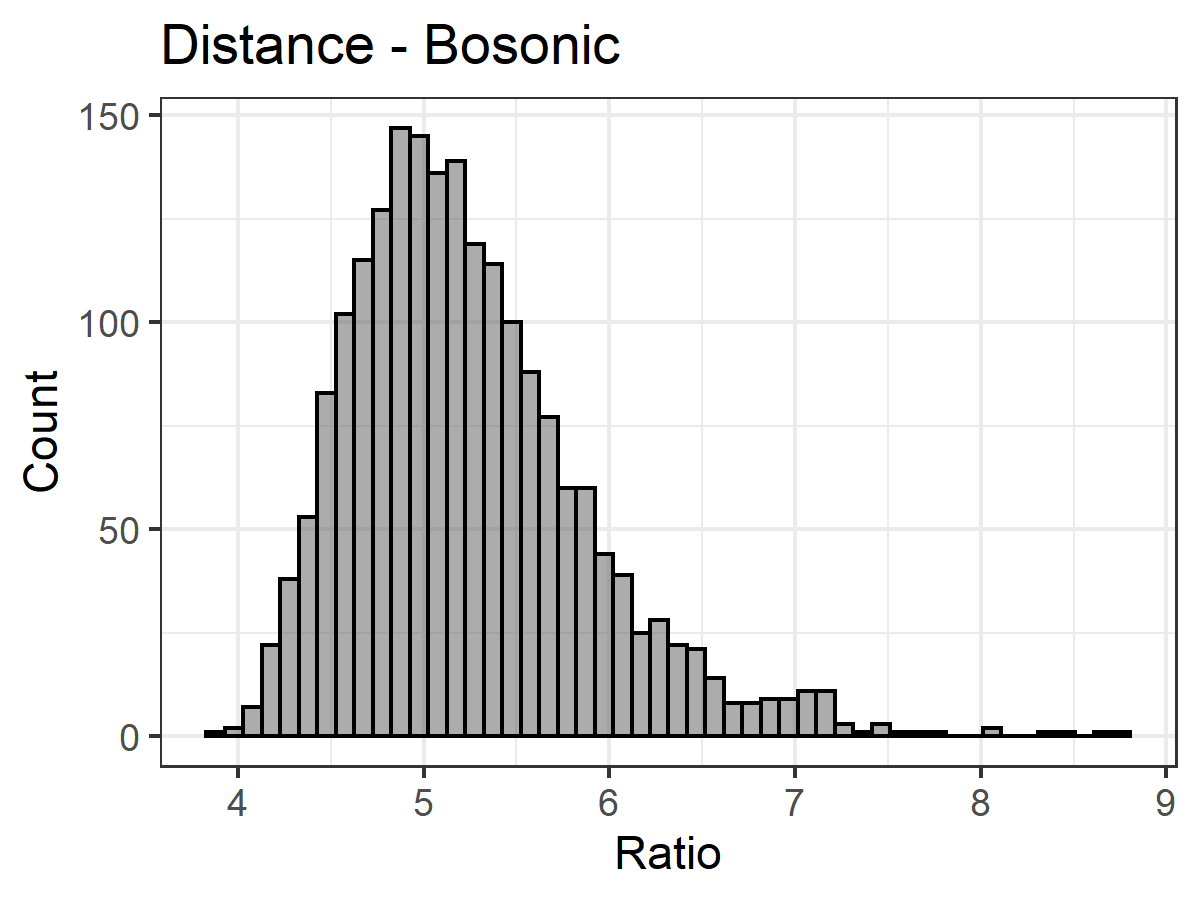}
	\caption{Random models of the matrix $A$. Plotted are the ratios $\frac{||A||_{\rm fer}^{\rho = 2, a}}{\left(||A||_{\rm fer}^{\rho = 1, a}\right)^2}$ (top row) and $\frac{||A||_{\rm bos}^{\rho = 2, a, \nu = 1}}{\left(||A||_{\rm bos}^{\rho = 1, a, \nu = 1}\right)^2}$ (bottom row), because this scales out any factors on the matrices. The sample size is 2000. Even though the scaling is removed, the ratio differs greatly from case to case, which suggests that there are many possible null modes just based on the symmetries.}
	\label{Fig:Random}
\end{figure}

I look at three random models for $A$:
\begin{itemize}
	\item Gaussian Hermitian (sometimes known as GUE \cite{Mehta})
	\item Real symmetric with positive entries uniformly distributed between 0 and 1
	\item A model where each entry is a random uniform number between 0 and the inverse euclidean distance in a 2D square grid.
\end{itemize}
These are treated purely numerically here, see Figure \ref{Fig:Random}, where I show the ratios
\begin{eqnarray}
	\frac{||A||_{\rm fer}^{\rho = 2, a}}{\left(||A||_{\rm fer}^{\rho = 1, a}\right)^2} \ &,&\ \frac{||A||_{\rm bos}^{\rho = 2, a, \nu = 1}}{\left(||A||_{\rm bos}^{\rho = 1, a, \nu = 1}\right)^2}\ .
\end{eqnarray}
This scales out any factors on the matrices as well as $\beta$. It can therefore act as a signature of a particular model type. Many configurations seem possible simply based on these 3 models.

\end{appendix}